# A TECHNICAL STUDY AND ANALYSIS ON FUZZY SIMILARITY BASED MODELS FOR TEXT CLASSIFICATION

Shalini Puri[1] and Sona Kaushik[2]

[1]M. Tech. Student, Birla Institute of Technology, Mesra, Ranchi, Jharkhand, India
eng.shalinipuri30@gmail.com
[2]M. Tech. Student, Birla Institute of Technology, Mesra, Ranchi, Jharkhand, India
sonakaushik22@gmail.com

## ABSTRACT

*In this new and current era of technology, advancements and techniques, efficient and effective text document classification is becoming a challenging and highly required area to capably categorize text documents into mutually exclusive categories. Fuzzy similarity provides a way to find the similarity of features among various documents.  In this paper, a technical review on various fuzzy similarity based models is given. These models are discussed and compared to frame out their use and necessity. A tour of different methodologies is provided which is based upon fuzzy similarity related concerns. It shows that how text and web documents are categorized efficiently into different categories. Various experimental results of these models are also discussed. The technical comparisons among each model's parameters are shown in the form of a 3-D chart. Such study and technical review provide a strong base of research work done on fuzzy similarity based text document categorization.*

## 1. INTRODUCTION

Text categorization [1] [2] is an upcoming and vital field in today's world which is most importantly required and demanded to efficiently categorize various text documents into different categories. Artificial Intelligence [3] – [5] provides many learning methods and paradigms to represent, interpret and acquire domain knowledge to help other documents in learning. Such categorization must produce the accurate and correct results with high performance. Due to the huge data size and complexity, data dimensionality reduction has also been a primary concern. Great levels of efforts have been put in this direction, so that the major problem of curse of dimensionality can be reduced.

Text documents clusterization [1] [2] [6] has been paid good attention. Many models and techniques have been developed for clustering. The clustering techniques can be applied to the web documents also. In this way, they can be categorized into their major and respective categories of business, stock, sports, cricket, movie, news and many more. Therefore, the unsupervised learning paradigm [6] is used to make the document clusters. It does not include any prior information and knowledge, that' why it requires complex text processing techniques.

Nowadays, text classification [7] – [16] [19] - [28] is gaining more attention and focus for text categorization activities [17] [18] even at the overhead of increased cost. Research is also being done for the fuzzy association, signature, c-means, algorithms and methods for categorization tasks. Text classification with fuzzy logic base provides a better forum to sufficiently categorize the text and web documents. It also results in justified solutions with reduced efforts. When it is combined with the feature clustering technique, it highly improves the representation of features. It further improves the storage performance and decreases the risks of feature ambiguity. Therefore, text classification techniques provide prior information and classification knowledge, so that classifiers can be made learnable to further categorize text and web documents. Many researchers are doing well in this area. Some of the applications in this field are, text classification system SECTCS (Smart English and Chinese Text Classification System) [8], segmenting handwritten text [9], nonlinear dimensionality reduction techniques [10] [11], complex linguistic features in context - sensitive text classification techniques [7] [12], cyber terrorism investigation [13], spam filtering [14] [15], topic spotting, email routing, language guessing, and many more. Text Classification and clustering are two opposite extremes with regard to the extent of human supervision they require. Real-life applications are considered somewhere in between, because unlabeled data is easy to collect but labelling data is more helpful.

As these techniques pay the attention on the accurate and correct categorization; they focus on the text pre-processing and document similarity analysis as well. During text pre-processing, the set of words are extracted to find out the concepts as features or words by using Verb-Argument Structures [6] or Pseudo Thesaurus [20]. In some research areas, bag-of-words [25] is found from the text documents. This word set is a huge collection of words that needs to be reduced further by using feature clustering [25] [28] methods. The resultant small collection of words is analyzed for the document similarity [16] [22] - [28]. If some of the documents are found similar, they are categorized into one. Many fuzzy similarity based models and algorithms have been introduced with the very nature of its membership functions [22] – [28], fuzzy association [24] [28], fuzzy C-means, production rules [19] [27]. Text classification using fuzzy based similarity is an essential task in today's categorization forum and typically, getting a great attention in various related application fields and areas. Nowadays, such concerns have been the part of many applications and related studies. Some of the applications are related to the learning evaluation [28] and education learning styles [19].

Section 2 discusses the key points and related aspects of theoretical background of fuzzy similarity based models and techniques. Section 3 discusses a technical comparative study on different fuzzy similarity based models. It discusses and shows various methods and their methodologies in detail. In section 4, an analytical discussion on the experimental results is given. Various results and their important concerns are discussed and shown with respect to different parameters. Finally, section 5 concludes the paper.

## 2. THEORETICAL BACKGROUND

Over the last decades, fuzzy similarity based text document classification has got attention very much and considered as an important research area. Different techniques, models and ways are searched to design a best categorization system. Such field is not only used in the small level organizations, industries and corporate, but also covers a vast community all around the world. The new techniques, their collaboration and research always open a new paradigm towards the advancements.

Current research studies show that fuzzy logic and its area of concerns provide efficient base for text categorization, dimensionality reduction, feature selection and extraction, and similarity analyzer related issues. Fuzzy logic is considered as a branch of logic especially designed for representing knowledge and human reasoning in such a way that it is amenable to processing by a computer [3]. The major concepts of fuzzy logic are fuzzy sets, linguistic variable, possibility distributions, and fuzzy if – then rules. Fuzziness or Degree of Uncertainty pertains to the uncertainty associated with a system, i.e., the fact that nothing can be predicted with exact precision. Practically, the values of variables are not always precise; rather approximate values are more likely to be known. The vagueness can adequately be handled using fuzzy set theory. This theory provides a strict mathematical framework using which vague conceptual phenomena can be studied rigorously. It is also called the property of language [3] – [5]. Its main source is the imprecision involved in defining and using symbols. It is a property of models, computational procedures, and languages. Hence, a fuzzy set is a collection of distinct elements with a varying degree of relevance or inclusion.

## 2.1. Feature Clustering

The concept of feature clustering [10] [11] [22] – [24] enhances the provision of text dimension criticality solution. It is an efficient way to compress the collected feature sets more, so that the resultant data can be handled and used properly without any loss. These clusters are represented either by the term of maximum frequency in a group (or cluster) [22] [24] or can be found by self constructing feature clustering algorithm [23]. Feature clustering is also done with the use of the pseudo-thesaurus by identifying each term [6] as noun, pronoun, adverb, adjective, delimiters etc. Researchers have shown that it helps to reduce the high dimensional data into smaller one adequately.

## 2.2. Fuzzy Association

Fuzzy sets pay an important and vital role in text categorization. They are widely recognized as many real world relations are intrinsically fuzzy. Fuzzy association [24] [28] is used to discover important associations between different sets of attribute values. A fuzzy association rule A $\Rightarrow$ C is very strong if both A $\Rightarrow$ C and C $\Rightarrow$ A are strong.

## 2.3. Fuzzy Production Rules

The novel method of rule-base construction and a rule weighting mechanism [19] [27] can result in a rule-base containing rules of different lengths, which is much more useful when dealing with high dimensional data sets.

## 2.4. Fuzzy Clustering and C-Means

In fuzzy clustering [28], each point has a degree of belonging to clusters, as in fuzzy logic, rather than belonging completely to one cluster. Thus, points on the edge of a cluster may be in the cluster to a lesser degree than points in the centre of cluster.

## 2.5. Fuzzy Signatures

Fuzzy signatures [26] are used in those applications and key areas which require the handling of complex structured data and interdependent feature problems. They can also used in special concerns where data is missing. So, this depicts many areas where objects with very complex and sometimes interdependent features are to be classified along with the evaluation of similarities and dissimilarities. This leads a complex decision model hard to construct effectively. Due to the very nature of fuzzy signatures of flexibility, it can be used for many text mining tasks, with the benefit of the hierarchical structuring; therefore, the text document classification models can be constructed [26].

## 3. A TECHNICAL COMPARATIVE STUDY ON DIFFERENT FUZZY SIMILARITY BASED MODELS

Research work on fuzzy similarity based models and techniques has taken a new turn for the text classification tasks with the involvement of different key concerns related to the fuzzy logic and sets. Therefore, these techniques provide better ways and solutions for categorization.

### 3.1. A Comparative Description on Various Proposed Techniques

The comparative detailed description on different techniques is described in table 1. It defines the challenges and problems occurred in each model, which are the related key issues. These models focus on different concerned issues and necessities of the text classification area. The similarity technique shows the efficient similarity criteria used in the model.

Table 1.  A Comparative Study among Various Fuzzy Similarity Based Models and Techniques.

| S N | Ref. No. | Problem Focused | Designed Aim | Similarity Technique |
|---|---|---|---|---|
| 1. | [22] | Comparative study of web-pages classification for Arabic Web-pages. | Arabic Web page classification using fuzzy similarity approach of fuzzy term relation category. | Fuzzy based similarity approach. |
| 2. | [23] | • Challenge of ambiguity in systems to handle natural language. • Issue of linguistic ambiguities found in text classification. | Proposed a text categorizer using Fuzzy Similarity methodology and Agglomerative Hierarchical Algorithms; Clique and Star, without needing to determine the number of initial categories. | Text categorizer of two algorithms based on fuzzy similarity based method. |
| 3. | [24] | The same word or vocabulary to describe different entities creates ambiguity, especially in the Web environment for large user population is large. | • A method of automatically classifying Web documents into a set of categories using the fuzzy association concept is proposed to avoid the ambiguity in word usage. | Similarity of distinct keywords of documents with the categories. |
| 4. | [25] | Need of a powerful method to reduce the dimensionality of feature vectors for text classification. | • Proposed a fuzzy similarity-based self-constructing algorithm for feature clustering. • Highly reduces the data dimensionality as each cluster, formed automatically, is characterized by a membership function with statistical mean and deviation. It chooses one extracted feature for each cluster. | Grouping of words in the feature vector of a document set into fuzzy clusters, based on similarity test. |
| 5. | [26] | Problem to identify the representation units as tokens using bag-of-words methods in some Asian Languages of non-segmented text. | • Proposed the fuzzy signature based solution using frequent max substring mining because of its language independency and favorable speed and store requirements. • Deals with cases to handle complex structure data, to handle overlapping information, | • Extracting index terms and use of a Super Substring definition to reduce the number of index terms. • Reduction in terms of finding out no super substring |
| | | | to include evolving information easily and to handle missing information. | pattern among index terms. |
| 6. | [27] | Challenge in high dimensional systems to generate every possible rule with respect to all antecedent combinations. | Proposed a method for rule generation, which can result in a rule-base containing rules of different lengths. | Production rule matching. |
| **Learning Evaluation** | | | | |
| 7. | [28] | Issues of expressing the fuzziness and uncertainty of domain knowledge and the semantic retrieval of fuzzy information. | • Produced an extended fuzzy ontology model/ <br> • Proposed a semantic query expansion technology to implement semantic information query based on the property values and the relationships of fuzzy concepts. | Semantic similarity and semantic correlation in fuzzy concept analysis. |

## 3.2. A Tour on Different Methodologies and Procedures

Various methodologies and procedures are depicted in table 2. These methodologies are shown in steps. [22], [23], [24], and [25] show that text documents or web documents are considered for text classification which use a predefined set of classes initially in the training phase. In [22], [23] and [24], the text is pre-processed and cleaned to extract all important features. In [25], a bag of words is used and processed to get the word patterns. Next, the fuzzy similarity techniques are applied as shown in table 1. Finally, text is classified using the classifier. Different methods have implemented different procedures to categorize the text.

The use of fuzzy signature for the text classification of the non-segmented text [26] shows that how the non-segmentable text can be segmented and classified. In [27], a rule based weighting technique is used to efficiently perform the data mining tasks. The learning evaluation using the extended fuzzy ontology model [28] is provided for learning techniques based classification. The given models have the key concern of the feature set reduction and improve the overall system performance.

Table 2.  Proposed Methodologies of Various Models and Techniques.

| Ref. No. | Description | Proposed Methodology |
|---|---|---|
| [22] | Web-pages Classification Methods using Fuzzy Operators Applied to Arabic Web-pages | 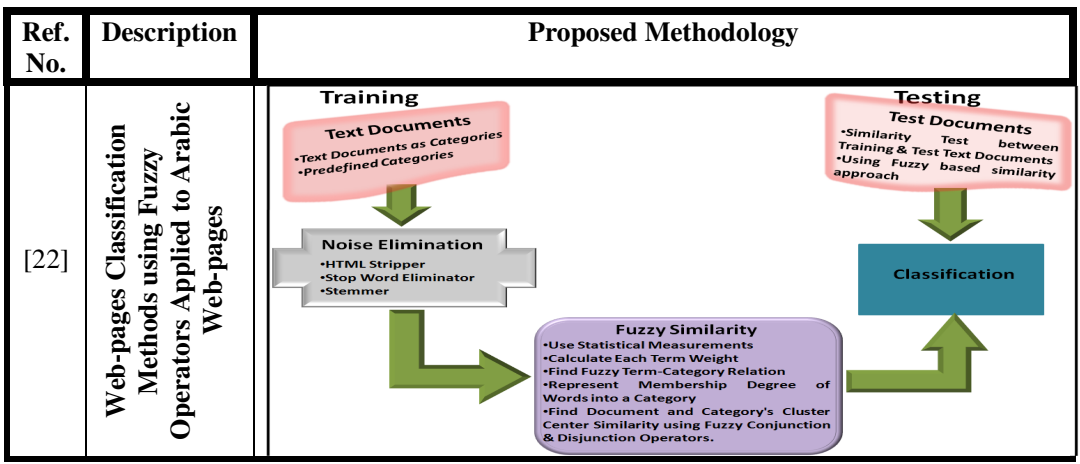 |

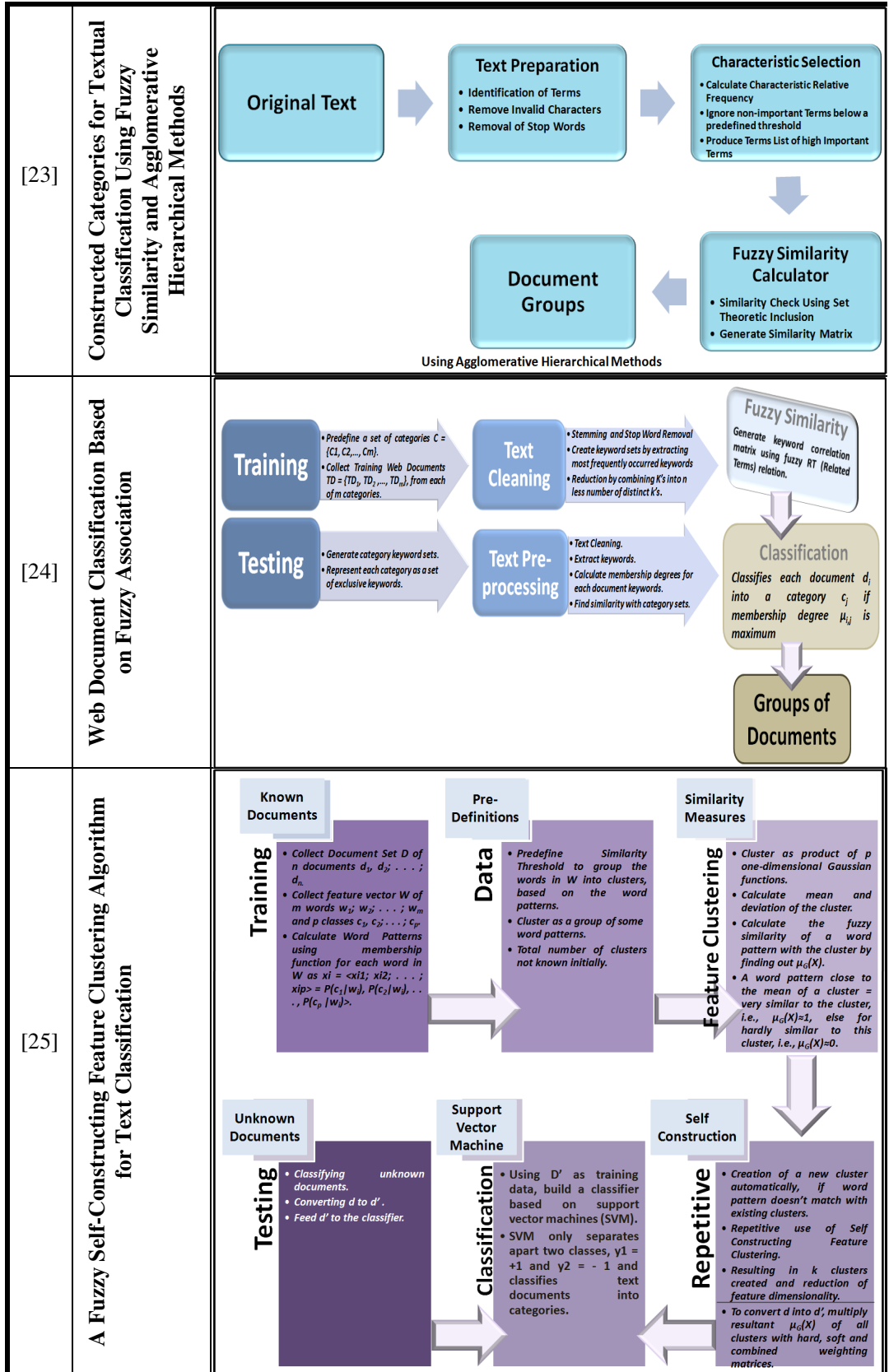

**[26]** — Exploring The Use of Fuzzy Signature for Text Mining

FMS Technique combined with Fuzzy Signature

- **Unit Representation**
  - To describe non-segmented text documents as vectors of index terms.
  - Using frequent max substring (FMS) mining to clearly define word boundaries.
  - Using a reduction rule, reduce total number of index terms & construct index structure corresponding to FMS having all frequent substrings, as,
    - user defined frequency threshold
    - by a super substring definition.

- **Fuzzy Signature**
  - A Nested Vector Structure or a Tree graph.
  - Provides semantic and logical connection of state variable.
  - Use of Fuzzy Aggregation to show relationship between lower and higher levels.
  - Result of parent signature= appropriate aggregation of their child signatures.

- **Categorizer**
  - Document classification and clustering

---

**[27]** — Efficient Fuzzy Rule Generation: A New Approach Using Data Mining Principles and Rule Weighting

- An M-class problem in an n-dimensional feature space.
- Use of m labeled patterns $Xp=[xp1, xp2, ..., xpn]$, $p=1, 2, ..., m$ from M classes.

→ To partition pattern space, use of only short fuzzy rules having a limited number of antecedent conditions are generated as candidate rules.

→ No information of an appropriate partitioning of each attribute, so adding the fuzzy set "don't care" to each attribute with membership function of defined as $\mu_{don't\ care}[x] =1$ for all values of x.
- Avoiding the exponential growth of the rule sets in each step.

→ Generate rules with 1 or 2 antecedents.
- Generate rules of higher dimensions.

→ Rule learning algorithm based on R.O.C analysis tunes the rule-base to have better classification ability.

---

**[28]** — Fuzzy Ontology Generation Model using Fuzzy Clustering for Learning Evaluation

Fuzzy Concept Analysis

- **Semantic Similarity**
  - Calculate $sem(c1, c2) = \alpha\ sem_{sub}(c1, c2) + (1-\alpha)\ sem_{con}(c1, c2)$, $\alpha \in [0,1]$
  - $sem_{sub}(c1, c2) = \theta\ sem_{sim}(c1, c2) + (1-\theta)\ sem_{relation}(c1, c2)$, $\theta \in [0,1]$
  - $sem_{sim}(c1, c2) = \frac{p(c1, c2)}{p(c1,c2)+p(T1,c2)+p(c1,T2)}$

- **Semantic Correlation**
  - $sem_{cor}(c1, c2) = w_{att}sim_{attribute}(c1, c2) + w_{rel}cor_{relation}(c1, c2)$
  - $cor_{relation}(c1, c2) = \frac{1}{n}\sum_{i=1}^{n} \frac{demote(v_{ci}, v_{ci})}{demote(v_{ci}, v_{ci})}$
  - $sim_{attribute}(c1, c2) = \frac{1}{n}\sum_{i=1}^{n} \frac{demote(v_{ci}, v_{ci})}{demote(v_{ci}, v_{ci})}$
  - $w_{att}+w_{rel} = 1$

Handle Uncertainty Information for Conceptual Clustering Using Fuzzy Concept Vocabulary Set

Expressing fuzzy concepts based on property values & semantic relationships

Relations → Fuzzy Concepts → Properties

Fuzzy Clustering
- Use all the concepts as nodes
- Link the concept node pairs for largest value of correlated degree according to the array A[MAXSIZE] for compete data.
- Set the threshold value $\delta \in [0,1]$.

- Fuzzy Ontology for Learning Evaluation
- Fuzzy Concept Set
- Property Set

## 4. AN ANALYTICAL DISCUSSION ON EXPERIMENTAL RESULTS

Various fuzzy similarity models for text classification have been successfully implemented. Their experimental results are shown and discussed in detail. The accuracy and performance parameters are evaluated and checked to see the utility of the methods and the current state - of – the - art.

### 4.1. Experimental Results: Data Sets and Evaluation

The experiments and results found for various models are discussed in table 3. It shows total data sets used, total number of categories generated and the results found for each technique. The data sets are considered from the newsgroups, newspapers, different text document pages of corpus, portals, Reuters, and repositories. Different categories are built initially in the training phase. These techniques have used documents from small corpus to large corpora, and considered few categories to many categories.

Experimental results found show that how the corresponding proposed technique is comparatively better than others. Some results have shown the performance and accuracy improvements, speed increase, reduced storage and many advantageous parameters.

Table 3. An Analysis Showing Different Experimental Results on Document Classification.

| Ref. No. | Data Set | Categories | Results Found |
|---|---|---|---|
| [22] | 50 *Arabic Pages*<br>5 Pages per Categories<br>6 Measures: Einstein, Algebraic, Hamacher, MinMax, Special case fuzzy and Bounded Difference. | 10 Categories: Autobiography (Auto), Children's Stories (Child), Economics (Eco), Health and Medicine (Hlth), Interviews (Intrv), Religion (Rlg), Science (Scnc), Short Stories (Short), Sociology (Socio), Tourist and Travel (Trst). | Accuracy Performance Achieved in the Decreasing order: Einstein bounded, Algebraic, ScFuzzy, Hamacher, MinMax. |
| [23] | Used *TeMario Corpus* of 100 texts<br>**Summaries**<br>• Manual Summaries<br>• Marked Manual Summaries<br>• Ideal Automatic Extracts<br>**Source Texts**<br>• With Title<br>• With out Title<br>• With Origin and Title | *Data Used for Simulation: From Origin and Title*<br>5 categories, each of 20 texts: from two Brazilian newspapers, *Folha de São Paulo* (Special, World, and Opinion) and *Jornal do Brasil* (Politics and International). | • A slight advantage of Clique algorithm over Star, but with greater number of groupings.<br>• Similar results of both for relationship rule.<br>• Excellent results for fuzzy similarity (set theoretic inclusion).<br>• Efficient technique of relative frequency in the characteristics selection phase. |
| | Data sets collected from 2 Web portals: *Yahoo! and Open Directory Project (ODP)* | *Yahoo! Portal 12 Categories* Arts & Humanities (art), Business & Economy (bus), Computers & Internet (com), Education (edu), Entertainment (et), | • Achieved higher accuracy in Fuzzy approach compared to the vector space model with Cosine coefficient.<br>• Total Accuracy |

| [24] | *Yahoo! : 350 most freque nt keywo rds from each categor y and total distinct keywo rds are 2033.* | *ODP: 350 most frequent keywords from each category and total distinct keywords are 1889.* | Government (gov), Health (health), News & Media (news), Recreation & Sports (rec), Science (sci), Social Science (sosci), Society & Culture (soc).<br><br>*ODP Portal 13 Categorie*s Arts (art), Business (bus), Computers (com), Games (game), Health (health), Home (home), Kids and Teens (kid), News (news), Recreation (rec), Science (sci), Shopping (shop), Society (soc), Sports (sport). | • Improvement in Fuzzy over Vector Method: Yahoo (TM): 13.7%, Yahoo (BM): 31.3%,<br>ODP (TM):17.7%, ODP (BM): 32%.<br>• For Accuracy Improvement of Vector Length of 10 in Yahoo!, TM: 17.9% and BM: 28.9%. |

| | • Used only English documents and ignorance of Non-English docs( World, Regional).<br>• Collected approximately 18,000 documents from each Web directory. | | | |

| Data Sets | Fuzzy Topmost | Fuzzy Bottommost | Vector Topmost | Vector Bottommost |
|---|---|---|---|---|
| Yahoo! | 81.5 | 60.1 | 67.8 | 28.8 |
| ODP | 84.8 | 78.1 | 67.1 | 46.1 |

| [25] | a. *20 Newsgroups Data Set,* about 20,000 articles taken from the Usenet newsgroups.<br>b. *Reuters Corpus Volume 1 (RCV1) Data Set,* 804,414 news stories.<br>c. *Cade12 Data* with skewed distribution and the three most popular classes represent more than 50 percent of all documents. | • In a, articles are evenly distributed over 20 classes, and each class has about 1,000 articles. Used two-thirds of the documents for training and the rest for testing.<br>• After preprocessing, found 25,718 features, or words, for this data set.<br>• In b, dividing the documents by the "LYRL2004" split into 23,149 training documents and 781,265 testing documents.<br>• There are 103 Topic categories and the distribution of the documents over the classes.<br>• In c, obtained a version of this data set, 40,983 documents in total with 122,607 features from which two-thirds, 27,322 documents, are split for training, and the remaining, 13,661 documents, for testing. | Proposed method runs faster and obtains better extracted features than other methods.<br>• In a, *for Execution time (sec.) of different methods on 20 Newsgroups data.* For 84 extracted features, only needs 17.68 seconds, but DC and IOC require 293.98 and 28,098.05 seconds.<br>• *Microaveraged Accuracy (Percent) of Different Methods*: S-FFC gets 98.46 percent in accuracy for 20 extracted features. H-FFC and M-FFC perform well in accuracy all the time, except for the case of 20 extracted features.<br>• *MicroP, MicroR, and MicroF1 (percent)*: S-FFC can get best results for MicroF1, followed by M-FFC, H-FFC, and DC.<br>• In b, proposed method runs much faster than DC and | |

| | | |
|---|---|---|
| | | IOC.<br>• H-FFC, SFFC, and M-FFC perform well in accuracy all the time.<br>In c, the proposed method runs much faster than DC and IOC. |
| [26] | • A corpus of 50 Thai Documents from Thai News Websites: 15 sport documents, 15 travel documents, 15 political documents and 5 education documents.<br>• Generated FMSs by frequent max substring technique from the document dataset<br>• Selection of 35 FSMs from document indexing. | *Sample of FMs extracted from 50 Documents*<br>• Competition, Athlete, Gold Medal, Semi final round, Sport type, Score, Competition result, Competition timetable, Thai travel exhibition, Tourist Attraction, The tourism authority of Thailand.<br>• *4 main Categories*: Sports, Travel, Political and Education.<br>• 2 Methods to recognize 4 document categories: Construct Fuzzy Signature with the use of membership function, construct 4 fuzzy signatures, one for each type of document, $A_{S(Sport)}$, $A_{S(Travel)}$, $A_{S(Political)}$, and $A_{S(Education)}$. | • With the use of fuzzy approach, no overlapping of the index terms occurred in the documents as in Self-Organizing Maps (SOM).<br>• Increased performance due to the use of Prior knowledge.<br>• Total number of FM in Sports: 8, Travel: 3, Political: 1 and Education: 0. *Competition* can be a part of Sports and Political.<br>• To recognize documents in both methods, fuzzy signature of FMSs is: $A_{S(Sport)} \rightarrow$ created by $\rightarrow$ Government, Education institution, non-profit organization, Business, $A_{S(Sport)} \rightarrow$ HTML keywords, $A_{S(Sport)} \rightarrow$ Inbound links $\rightarrow$ Quantity, Categories, $A_{S(Sport)} \rightarrow$ from 35 FMSs. |
| [27] | *Used a different of UCL ML repository data sets*<br>• Generated all the rules of length 1, 2, 3, and 4 (i.e. having 1, 2, 3, and 4 number of antecedent conditions excluding don't cares).<br>• Used 10CV technique: Case of n - fold cross validation.<br>• Construction of a rule-base by selecting 100 candidate rules from each class | *Some statistics of the data sets used in computer simulations*<br><br>**Statistics table:**<br>(see below)<br><br>**Classifier comparison table:**<br>(see below) | • Improves classification accuracy by considering cooperation in a rule-base tuned by rule weighting process.<br>• Increasing the maximum length of rules in the initial rule-base improves the classification accuracy.<br>• Comparing proposed classifier and best case of C4.5: Improvement of 0.7, 0.3, 5.5 and 3.3 in first 4 cases of the proposed classifier, but decreased accuracy of 4.5 in 5th one. |

*Some statistics of the data sets used in computer simulations*

| Data set | No. of Attribute | No. of Patterns | No. of Classes |
|---|---|---|---|
| Iris | 4 | 150 | 3 |
| Wine | 13 | 178 | 3 |
| Thyroid | 5 | 215 | 3 |
| Sonar | 60 | 208 | 2 |
| Bupa | 6 | 345 | 2 |
| Pima | 8 | 768 | 2 |
| Glass | 9 | 214 | 6 |

| Data Sets | Proposed Classifier | C4.5 Classifier | |
|---|---|---|---|
| | | Worst | Best |
| Iris | 95.6 | 94 | 94.9 |
| Pima | 7.3 | 72.8 | 75 |
| Sonar | 82.2 | 67.4 | 76. |

| | | | | | 7 |
|---|---|---|---|---|---|
| | using the selection metric. | | Wine | 97.7 | 92.2 | 94.4 |
| | | | Glass | 68.2 | 68.8 | 72.7 |

| [28] | *Learning Evaluation for Teaching Field*<br><br>Consider Entity concept *"student"*<br>a. *Property Set:* {learning attitude, learning ability, text scores,…}<br>b. *Property Value set:*<br>• learning attitude ( very good, basic good, bad, very bad, …).<br>• learning ability (most strong, very strong, strong, general weak, weak, great weak, …).<br>• text scores (extremely high, high, medium, slight low, low, …). | • A set of categories {C1, C2,…, C7} ⊆ {Concept Vocabulary Set}.<br>• *Concept Vocabulary Set Values*: {excellent, good, bad, medium, strong, high, low} and the semantic relationship of every concept pair.<br>• *Predefinitions:* factor $\alpha = 0.5$, $\Theta = 0$, $w_{att} = 0$, and Threshold Value $\delta = 0.9$.<br>• *Calculations:*<br>$sem_{sim}(c1, c2) = sim_{heuristic}(c1, c2) = 0.9$<br>$sem_{corr}(c1,c2) = corr_{relation}(c1, c2) = 1$<br>$sem(c1,c2) = 0.95$ | • Production of Concept Connected Graph with total 28 Different entries in 7*7 matrix of C1 to C7, where the result found (without duplication of entries) as, total number of 0 is 8 times, 1 is 7 times, 0.5 is 4 times, 0.95 is 3 times, 0.8 is 3 times, 0.35 is 2 times, and 0.9 is 1 time.<br>• Graph Flow and connections among concepts are, C1→C6, C5, C2, and C2→C4→C7→C3.<br>• Performance Evaluation based on Precision.<br>• Determining the relevance of the information and obtaining the exact information.<br>• Shows better results found in extended fuzzy ontology model than Classical Ontology Method. |

## 4.2. Various Experimental Results on the Models

The experimental results of various models show their good performance and accuracy concerns. In figure 1, these models are discussed and their studies, results, comparisons of experimental results are shown. The bars in chart are individual and independent in their identity. These results are not compared with each other; they only provide their data, and respective details.

Fuzzy term-category relation [22] is shown by manipulating membership degree for the training data and the degree value for a test web page. Six measures are used and compared where the best performance was achieved by Einstein. Accuracy performance of these algorithms in the decreasing order is shown in figure 1. With this, the training data collected from different sources is normalized and pre-processed and then these measures are applied on it. Text categorization based on the Agglomerative Hierarchical Methodology [23] with the use of fuzzy logic. As for the use of the star and clique algorithms used in the agglomerative hierarchical methodology to identify the groups of text by specifying some type of relationship rule, they obtained similar results, but the clique algorithm showed a slight advantage when compared to the star, despite having created greater number of groupings. In figure 1, star and clique algorithms are compared for the parameters, number of categories, group of 10 or more texts and categories of only one text. Clique shows better results than star.

To automatically classify the web documents using the fuzzy association concept [24], the relationship is captured among different index terms in documents. This approach is compared with vector space model approach and it shows improved results than VSM. To see the effect of different keyword selections for category vectors, 2 different alternatives are there: Selecting from the most frequently occurred keywords(topmost) and selecting from the least frequently occurred keywords (bottommost) with varying vector lengths have been used. Yahoo! And ODP portals are compared with each other for the topmost and bottommost cases as shown in figure 1. In [25], a small part of the total result is shown in chart. It is only shown for a subsection of the 20 newsgroups.

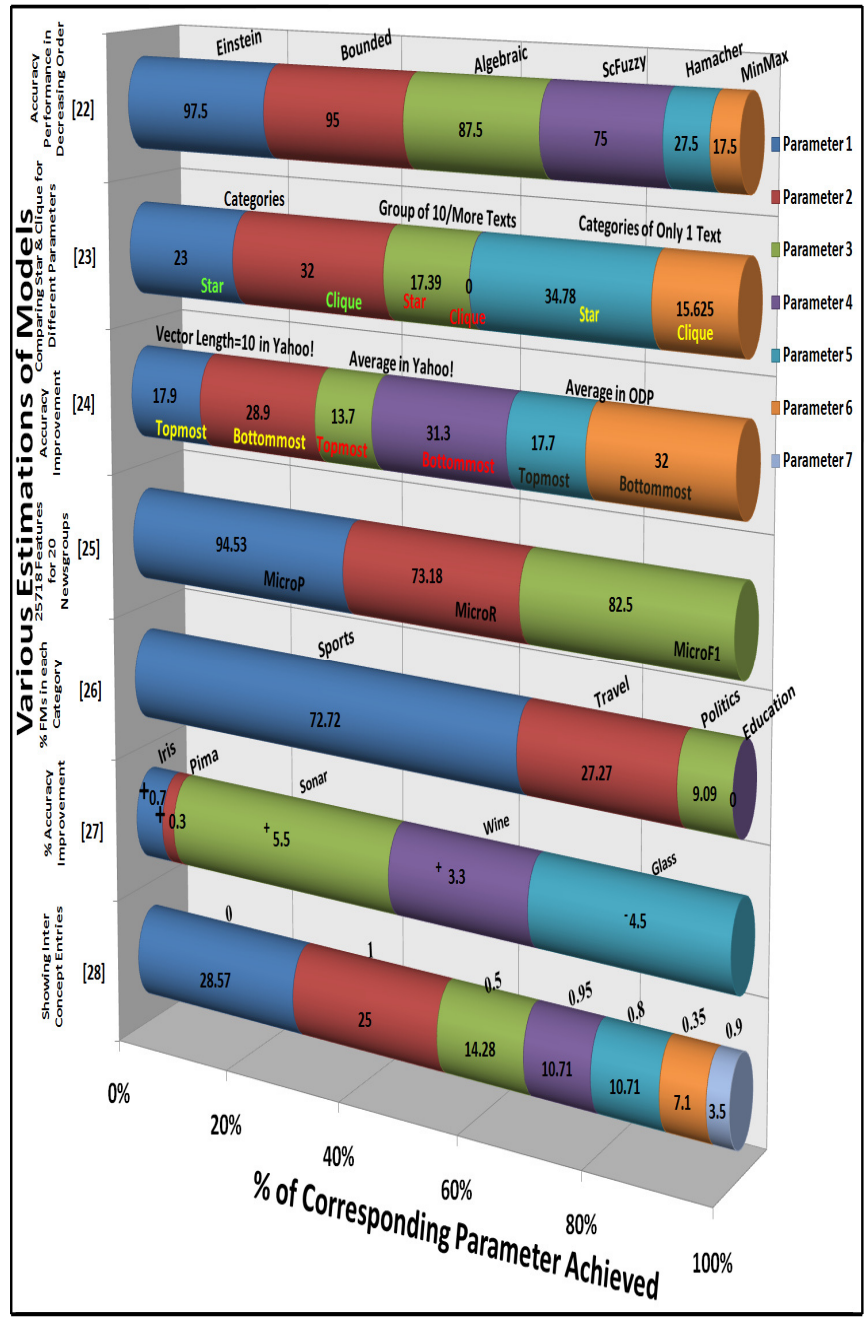

Figure 1.  Various Models, their Parametric analysis with Experimental Results

[26] discusses the simple category distribution in each of the 4 type of documents of sample data. In [27] and [28], the parameters are calculated as given in the table 3. In [27], Iris showed better results over others. In [28], to make the information semantization and to improve the accuracy of information retrieval, it adopted a fuzzy concept semantic analysis for clustering to generate learning evaluation ontology. It achieves high information retrieval and improves efficiency as compared to fuzzy ontology.

# 5. CONCLUSIONS

In this paper, different fuzzy similarity related algorithms and methodologies are discussed in detail. Different researches depict good results with the underlying techniques, mechanisms and methodologies. The experimental results provide good fuzzy based text classification with high accuracy. These models focus on new kinds of different classification issues and techniques. Therefore, these research studies and their survey contribute in providing the information about advanced fuzzy classification, related models and techniques.

The analytical review provides a simple summary of the sources in an organizational pattern and combines both summary and synthesis to give a new interpretation of old material. Therefore, it aims to review the critical points of current knowledge of research work including substantive findings as well as theoretical and methodological contributions. Additionally, their experimental results and their parametric data are sufficiently described and compared independently. Such comparative studied and technical analysis charts provide a strong base to understand the use of fuzzy and its related concerns. Various experimental results have proven themselves good for the models and techniques. The utility of fuzzy logic and its areas give a good effect on text mining and text classification. Therefore, fuzzy similarity is used in many application areas and fields all around the world for categorization.

## ACKNOWLEDGEMENTS

We would like to give our special thanks to Asst. Prof. Pankaj Gupta, Dept. of Computer Science, Birla Institute of Technology, Noida Extension Centre, Uttar Pradesh, India and Dr. Vikas Saxena, Dept. of Computer Science, Jaypee Institute of Information Technology, Noida, Uttar Pradesh, India for their help and guidance.

**Authors**

Shalini Puri received the B. E. Degree in Computer Science from Mody College of Engineering and Technology, Sikar, Rajasthan, India in 2002. She is pursuing M. Tech. in Computer Science at Birla Institute of Technology, Mesra, Ranchi, Jkarkhand, India. She is currently working as an Assistant Professor in a reputed engineering college in India. She has published many international journals and presented papers in IEEE conferences. Her research areas include Artificial Intelligence, Data Mining, Soft Computing, Graph Theory, and Software Engineering.

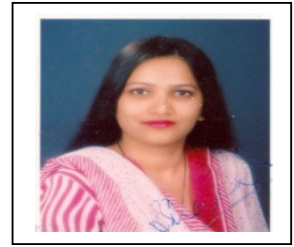

Sona Kaushik belongs to National Capital Region, New Delhi, India. She received the B.Tech. Degree in Information Technology in 2007. She is currently working as a System Engineer in a reputed IT Organisation and pursuing Masters in Technology from Birla Institute of Technology, Mesra, Ranchi, India. She has published many international journals and presented papers in IEEE conferences. Her research interests are in information security, network security, security engineering and cryptography.

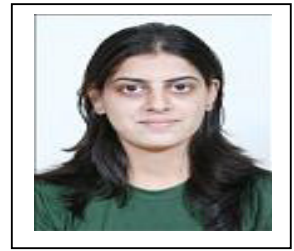